# High pressure operation of the Photon-Assisted Cascaded Electron Multiplier

F. D. Amaro, J. F. C. A. Veloso, J. M. F. dos Santos, A. Breskin, R. Chechik, A. Lyashenko

*Abstract–* We present the operation of the recently introduced Photon Assisted Cascaded Electron Multiplier (PACEM) in xenon at high pressure. The PACEM is a multi step electron multiplier where the VUV scintillation produced in the electron avalanches is used for signal propagation: the VUV scintillation produced in the first element of the cascade induces the emission of photoelectrons from a CsI photocathode placed on the top-surface of the second element. These photoelectrons are further multiplied, via charge avalanche. A metallic mesh electrode placed between the first and the second elements of the cascade completely blocks the charge transfer between them. Optical gains of $10^3$ were achieved in xenon at atmospheric pressure, dropping to 25 at 3.3 bar, for applied voltages of 700 and 1100 V, respectively. Taking into account the subsequent charge multiplication, total gains are higher than those obtained with a triple GEM and double THGEM.

## I. INTRODUCTION

The Photon-Assisted Cascaded Electron Multiplier (PACEM) [1] was developed as an alternative for blocking avalanche-induced ions in cascaded gaseous electron multipliers. The PACEM is a two step cascade multiplier with a Micro Hole and Strip Plate (MHSP) [2] as the first element (fig. 1). A metallic grid, with high optical transparency, is placed between the first and second elements of the cascade, electrically isolating one from the other. The signal is transmitted from the first element to the next one using the VUV scintillation produced during the electron avalanches that take place in the first element. A CsI reflective photocathode is deposited on the top of the second element, promoting the extraction of photoelectrons from its surface. The number of UV-induced photoelectrons extracted from the CsI photocathode depends on the UV photons wavelength [3]. The photoelectrons extracted from the reflective photocathode are further multiplied in the second element of the cascade by charge avalanche mechanisms. This second element may be any suitable gaseous electron multiplier (e.g. a GEM [4], fig.1).

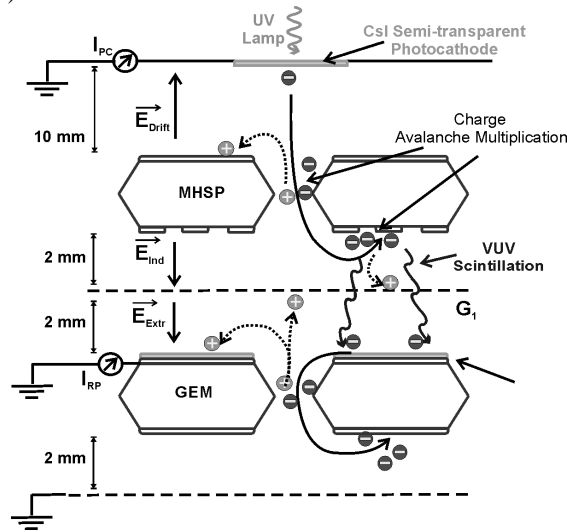

Fig. 1. Schematic view of the PACEM detector on a MHSP/GEM configuration.

The optical signal transmission is only viable in highly scintillating gases such as heavy noble gases (Ar, Kr and Xe) [1][5][6] and $CF_4$ [7]. The PACEM principle was first validated in pure xenon, demonstrating the viability of the VUV scintillation-signal transmission and the high efficiency of the ion blocking grid [1][6]. In $CF_4$, the higher photoelectron extraction efficiency from the CsI photocathode into the gas compensated for the lower scintillation and photoelectron yields, reaching similar performance than that obtained in xenon [7].

High pressure xenon detectors are used for hard x-ray and gamma ray imaging, in medical and astrophysics applications

Manuscript received July 30, 2008. This work was supported in part by the Project POCI/FP/63962/2005 through FEDER and FCT (Lisbon) programs, by the MINERVA Foundation and by the Israel Science Foundation, grant No 402/05. A. Breskin is with the W.P. Reuther Professor of Research in the Peaceful Use of Atomic Energy

F. D. Amaro is with the Physics Dept., University of Coimbra, Coimbra, Portugal (telephone: 351-239410667, e-mail: famaro@gian.fis.uc.pt).

J. F. C. A. Veloso is with the Physics Dept., University of Aveiro, Aveiro, Portugal (telephone: 351-234378108, e-mail: jveloso@fis.ua.pt).

J. M. F. dos Santos is with the Physics Dept., University of Coimbra, Coimbra, Portugal (telephone: 351-239410667, e-mail: jmf@gian.fis.uc.pt).

A. Breskin is with the Dept. of Particle Physics, The Weizmann Institute of Science, Rehovot, Israel (telephone: 972-8-9342645, e-mail: amos.breskin@weizmann.ac.il).

R. Chechik is with the Dept. of Particle Physics, The Weizmann Institute of Science, Rehovot, Israel (telephone: 972-8-9342645, e-mail: rachel.chechik@weizmann.ac.il )

A. Lyashenko is with the Dept. of Particle Physics, The Weizmann Institute of Science, Rehovot, Israel (telephone: 00972-8-9342064, e-mail: alexey.lyashenko@weizmann.ac.il )

[8], and rare-event detection such as double-beta decay [9] and dark-matter search [10]. Charge readout devices integrated in the detector volume are desirable. Recent developments in micro-pattern detectors show the possibility of operating gas electron multipliers at high pressures, also in combination with CsI photocathodes [11][12][13]. The PACEM could present an alternative solution for such applications.

The gain of the PACEM is dependent on the number of electrons produced by charge multiplication on the MHSP and on the gas scintillation yield, i.e., the number of photons produced per electron in the gas when the electrons are accelerated towards the anode strips. On the other hand, another important factor is the UV-induced photoelectron transfer efficiency from the CsI photocathode into the GEM; it largely depends on the return of the photoelectrons to the photocathode due to backscattering in the gas, which also depends on the gas pressure.

It is therefore necessary to determine the behavior of the PACEM detector at high pressure, since the fast drop in charge gain at the MHSP could be compensated by a slower reduction of the scintillation yield and of the photoelectron transfer efficiency with increasing pressure.

In the present work, the operation of the PACEM in pure xenon at pressures up to 3.3 bar was investigated. The optical gain, i.e., the number of photoelectrons per primary electron extracted from the reflective photocathode and the relative IBF, i.e., the number of ions flowing back to the drift region per primary electron, were studied as a function of the applied voltages and xenon pressures.

## II. EXPERIMENTAL METHOD AND SETUP

The PACEM used in this work is presented in fig.1; it is described in details in refs.[1] and [6]. The MHSP and the GEM share the same production technique being both made from a micro-perforated 50 μm thick Kapton® foil with a $28 * 28$ mm$^2$ active area, covered on both sides with a gold-plated 5 μm copper layer. The MHSP has an additional strip pattern on its bottom face (fig. 2): 100 μm wide cathode strips, with 70 μm diameter biconical holes aligned along their centers, and 20 μm wide anode strips etched between the cathode strips. With this arrangement the MHSP has two independent electron-multiplication regions: inside the holes, established by the voltage difference between the top electrode and the cathode strips, $V_{CT}$, and at the anode strips, established by the voltage difference between cathode and anode strips, $V_{AC}$. Compared to the GEM, the MHSP allows for higher scintillation and charge gains, [6][11]. Since the MHSP and GEM present the same top electrode configuration, the primary electron collection efficiency into the MHSP holes is similar as that of the GEM, as measured in [14]. In addition, depending on the electric field between the MHSP bottom face and the grid electrode ($G_1$, fig. 1), only ~20% of the anode-avalanche ions drift back through the MHSP holes [15].

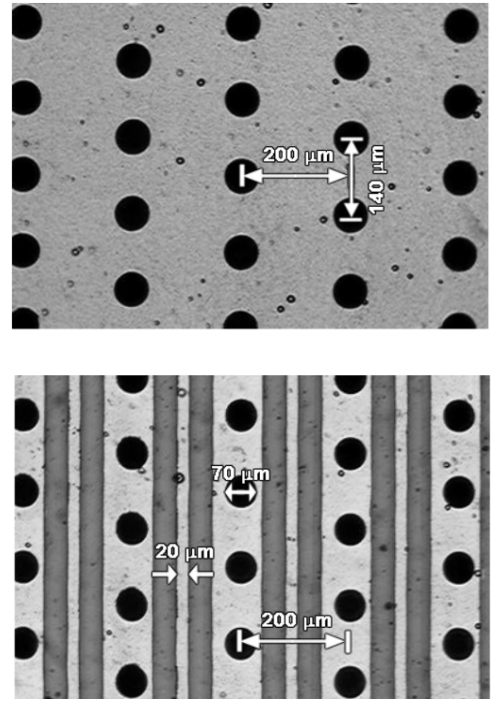

Fig. 2. Top (upper view) and bottom (lower view) side of the MHSP.

The scintillation produced during the electron avalanches of the MHSP induces the extraction of photoelectrons from the reflective CsI photocathode (about 2500 Å thick) deposited on the top of the GEM, fig.1. The stainless-steel grid (80 μm diameter wires, 900 μm spacing), $G_1$, is connected to ground potential, blocking any charge transfer (electrons and ions) between the MHSP and the GEM and vice-versa. The photoelectrons emitted by the CsI photocathode are transferred into the GEM holes, where they undergo multiplication. The avalanche electrons extracted from the holes are collected on the GEM bottom electrode, or may follow on to subsequent multiplier elements of a larger cascade and/or to a charge readout device.

For the measurements presented in this work only the photoelectron current emitted by the reflective photocathode was recorded; there was no further photoelectron multiplication at the GEM holes. The grid $G_1$ was polarized while the top GEM electrode was kept at ground potential resulting in an extraction field of 1.0 kV/cm at 1 bar. This field was increased proportionally to the pressure, resulting in a constant field of 1.0 kV/cm*bar.

The detector was filled with xenon and operated in sealed mode. Before filling the detector, it was pumped down to $10^{-6}$ mbar and then filled with the required pressure, up to 3.3 bar. During the operation, the xenon was continuously purified using non-evaporable getters, SAES St707, placed in a vessel connected to the detector. All the electrodes of the detector were independently biased using CAEN N741A power-supply units. The present studies were performed operating the PACEM in current mode, recording the current on the electrodes with Keithley 610C electrometers.

The primary-electron current, $I_{PC_0}$, was generated from a CsI semitransparent photocathode placed 10 mm above the first element, as shown in fig.1. This photocathode, 250 Å thick, was deposited on a 5 mm thick quartz window and was irradiated by a UV Hg(Ar) lamp. The photocathode was connected to ground potential and the photoelectrons were extracted into the gas by the electric field applied across the drift region. Measurements were performed using drift electric fields of 0.1 and 0.5 kV/cm*bar, ensuring full primary electron collection efficiency [16].

The optical gain was defined as the ratio between the number of photoelectrons extracted from the reflective photocathode, due to the gas scintillation produced in the MHSP electron avalanches, and the number of primary electrons extracted from the semitransparent photocathode [6]. Due to the geometry of the detector and to the optical transmission of the MHSP and of the blocking grid (12% and 84% respectively), part of the UV photons emitted by the Hg (Ar) lamp reached the reflective photocathode placed on the top of the GEM (see fig. 1). The photoelectrons extracted from the reflective photocathode due to the direct irradiation of the UV lamp yielded a contribution, $I_{RP_0}$, to the total photoelectron current measured on the reflective photocathode, $I_{RP}$.

The optical gain was calculated according to:

$$Optical\ Gain = \frac{I_{RP} - I_{RP_0}}{I_{PC_0}} \quad (1).$$

Both $I_{RP_0}$ and $I_{PC_0}$ were measured for null electric fields in the MHSP, i.e. $V_{CT} = V_{AC} = 0$ V, and in the GEM, i.e. $\Delta V_{GEM} = 0$ V, and for an extraction field of 1.0 kV/cm*bar in the region between the reflective photocathode and the wire mesh $G_1$.

In addition, the charge gain in the MHSP could be estimated by adding the currents on the anode and cathode of the MHSP. These currents were measured using the CAEN N471A power supplies current monitors.

$$Charge\ Gain = \frac{I_{Catode} + I_{Anode}}{I_{PC_0}} \quad (2).$$

The currents measured at the semi-transparent photocathode could also be used to evaluate the ion back-flow (IBF) of the positive ions, per primary electron, into the drift region:

$$\frac{IBF}{pe} = \frac{I_{PC} - I_{PC_0}}{I_{PC_0}} \quad (3).$$

where $I_{PC}$, is the current recorded at the semitransparent photocathode and $I_{PC_0}$ is the same as above. These ions flow back into the conversion/drift region of the detector where they pose limitations to its operation. In particular, the effects of the ions in gaseous detectors are of particular relevance for the development of Time Projection Chambers (TPC) [17] and visible sensitive Gaseous Photomultipliers (GPM) [18]. In the PACEM the number of ions that flow back to the drift region is independent of the total gain of the cascaded multiplier. Since the PACEM has the first element electrically decoupled from the others, only a fraction of the ions produced on the first element of the cascade contributes to the IBF. The gain on the next elements can be increased without causing any further contribution to the IBF [6].

III. EXPERIMENTAL RESULTS AND DISCUSSION

The optical gain and absolute IBF of the PACEM multiplier operating in xenon were measured for pressures up to 3.3 bar as a function of the total voltage, $V_{Total} = V_{CT} + V_{AC}$, applied on the MHSP. For each series of measurements the voltage across the holes of the MHSP, $V_{CT}$, was raised up to its maximum value, below the onset of discharges, keeping the electric field between cathode and anode strips null ($V_{AC} = 0$; the so-called "GEM mode" operation of the MHSP). Once the $V_{CT}$ voltage was set at its maximum value, $V_{AC}$ was raised, resulting firstly on charge transfer, from the cathode to the anode, and then, for higher $V_{AC}$ values, on charge multiplication at the anode strips. The results obtained are presented in figs. 3 and 4.

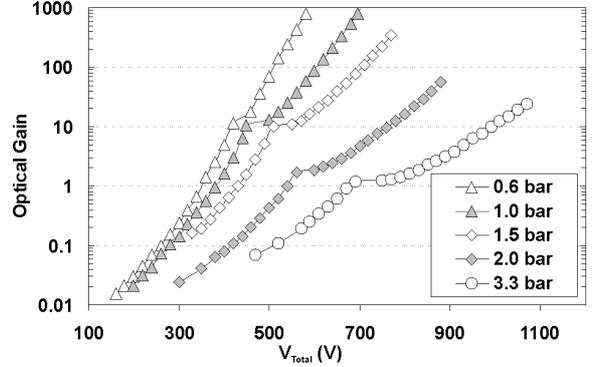

Fig.3. Optical gain of the PACEM detector in Xe, as a function of $V_{Total}$, for pressure up to 3.3 bar and for $E_{Drift} = 0.1$ kV/cm*bar.

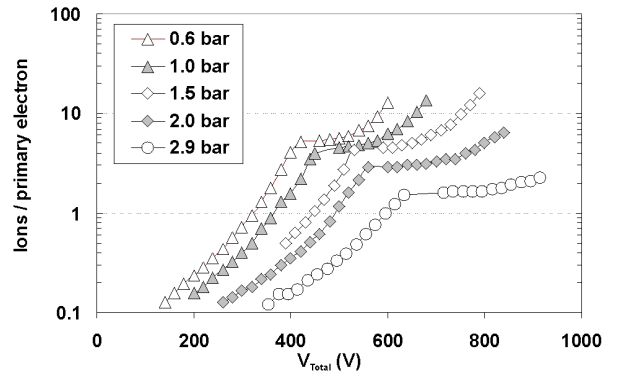

Fig. 4. Absolute IBF of the PACEM detector in Xe, as a function of $V_{Total}$, for pressure up to 2.9 bar and for $E_{Drift} = 0.1$ kV/cm*bar

For both the optical gain and absolute IBF the inflection points in the curves of figs. 3 and 4 represent the condition where $V_{CT}$ reaches its maximum value and where $V_{AC}$ starts being raised. After this point the optical gain and absolute IBF behave differently; while the optical gain increases steadily with increasing $V_{Total}$, the absolute IBF shows a slower increase with increasing $V_{Total}$. This is due to the lower threshold for scintillation relatively to ionization in xenon. For low values of $V_{AC}$ the electric field in the region between anode and cathode is too low to induce ionization of the xenon atoms. Therefore, the number of ions produced only approaches an exponential increase for higher values of $V_{AC}$, when charge multiplication takes place between anode and cathode. The scintillation yield is enhanced for low values of $V_{AC}$ in comparison to the production of ions.

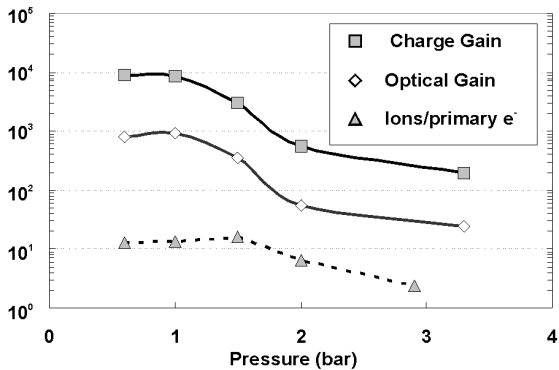

Fig. 5. Maximal values of the charge, optical gain and respective relative IBF as a function of the pressure. The values presented were measured for $E_{Drift} = 0.1$ kV/cm*bar.

A maximum optical gain of 940 was recorded at 1 bar. This gain was reduced to 56 and 25 at 2.0 and 3.3 bar, respectively. The number of ions per primary electron reaching the drift region, at maximum optical gain, dropped from 13.5 at 1 bar to 6 and 2.3 at 2.0 and 2.9 bar, respectively (fig.4). Even for the maximum optical gain, the obtained IBF value was low, close to $10^{-3}$ for total cascade gains of $10^4$ at 1 bar. For higher pressures the IBF value approached the $10^{-4}$ limit, required for high-rate TPCs [19]. These values are much lower than those obtained for triple-GEM cascades [20] and similar to that recently reached in MHSP/GEM configurations.

Fig. 5 summarizes the maximum optical gain and the maximum MHSP charge gain, achieved before the onset of discharges. Similar behavior was measured for the maximum optical and charge gains of the MHSP, as the scintillation depends on the electrons produced in the electron avalanche. This behavior is similar to the one observed in a previous work [11]. The IBF value obtained at the maximum gain is also depicted in fig.5.

In fig. 6 we present the ratio between charge and optical gain of the PACEM detector as a function of the optical gain, for different pressures. For low values of optical gain, corresponding to the situation where $V_{AC} = 0$ V (the "GEM mode" operation of the MHSP), the ratio optical-to-charge gain decreases, revealing that the charge production is favored relatively to the scintillation within the holes of the MHSP as the electric field increases. Similar observation was done in [21], when measuring the GEM scintillation in xenon using an avalanche photodiode. As $V_{AC}$ increases, the ratio optical/charge gain increases steadily until it reaches a value close to 0.1, almost independently from pressure.

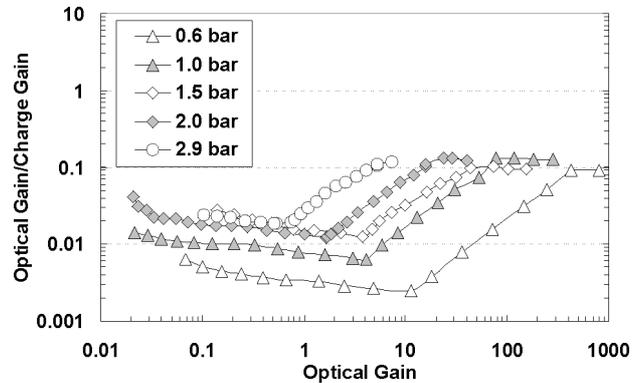

Fig. 6. Ratio between optical and charge gain as a function of optical gain (by increasing $V_{Total}$), for pressure up to 2.9 bar and for $E_{Drift} = 0.1$ kV/cm.

We note that, in spite of the ten-fold reduction of optical vs. charge gain in the MHSP, the PACEM has an additional avalanche-gain in the last element (a GEM in the present configuration, fig.1). Therefore, the total charge gain of the PACEM has to include the charge gain of this element [22]. This should result in charge gains of the PACEM superior to that of the triple-GEM [12]. In addition, cascading two or more PACEM elements (i.e. a cascade of MHSP and mesh sets) should result in higher total gains while maintaining the total voltage applied to the cascade unchanged, in opposition to the operation of GEM and Thick-GEM (THGEM) cascades [12][13]. The voltages applied to the successive MHSPs/mesh sets in cascade will be exactly the same.

## IV. CONCLUSIONS

We have demonstrated an efficient operation of the PACEM at xenon pressures up to 3.3 bar. Maximum optical gains of $10^3$, 56 and 25 were obtained for xenon pressures of 1, 2 and 3 bar, respectively. Taking into account the charge gain of the final PACEM element (not part in this work), the overall gain of the PACEM is expected to be higher than that reached with triple-GEM cascades, at lower applied voltages. Cascading several PACEMs would result in much larger overall gains without increasing the total voltage of the cascade, e.g. in opposition to GEM and THGEM [23] cascades. This could present an benefit for high-pressure operation.

The charge gain of THGEMs is similar or higher than that of the MHSP and these devices are more robust and simple to operate, when compared to the latter and to the GEM. Therefore, a configuration of THGEM/mesh/THGEM could provide higher gains compared to the present PACEM configuration; this could be an asset for many applications.